\documentclass{article}
\usepackage{xcolor}
\usepackage{hyperref}
\usepackage{graphicx} 
\usepackage{authblk}  
\usepackage{amsmath} 
\usepackage{booktabs} 
\usepackage{amssymb}
\usepackage{pifont}
\usepackage{makecell}
\usepackage{multicol}

\usepackage[most]{tcolorbox}
\newcommand{\example}[3]{\begin{tcolorbox}[ colframe=blue!60!yellow, colback=blue!5!white, title={Example Problem}]
\textbf{#1}

\vspace{0.1in}
{#2}

\vspace{0.1in}
\textbf{Problem Type}: {#3}
\end{tcolorbox}
}

\newcommand{\analysis}[4]{\begin{tcolorbox}[ colframe=red!50!yellow, colback=blue!5!white, title={Model Behavior Analysis}]
\textbf{Problem: #1}

\vspace{0.1in}
{#2}

\vspace{0.1in}
\textbf{Model Output Excerpts}: {#3}

\vspace{0.1in}
\textbf{Analysis}: {#4}
\end{tcolorbox}
}

\newcommand{\cross}{\ding{55}}
\newcommand{\halfcheckmark}{\checkmark\textsuperscript{\raisebox{0.2ex}{\kern-0.48em\scriptsize\cross}}}

\newtcolorbox{poscarbox}{
  colback=gray!5,
  colframe=gray!40,
  boxrule=0.3pt,
  arc=1pt,
  left=4pt,right=4pt,top=2pt,bottom=4pt,
  fontupper=\tiny\ttfamily,
  breakable
}

\title{QMBench: A Research Level Benchmark for Quantum Materials Research}

\author[1]{Yanzhen Wang}
\author[2]{Yiyang Jiang}
\author[3]{Diana Golovanova}
\author[3]{Kamal Das}
\author[3]{Hyeonhu Bae}
\author[3]{Yufei Zhao}
\author[2]{Huu-Thong Le}
\author[2]{Abhinava Chatterjee}
\author[2]{Yunzhe Liu}
\author[2]{Chao-Xing Liu}
\author[1]{Felipe H. da Jornada}
\author[2,3]{Binghai Yan\thanks{binghai.yan@psu.edu}}
\author[4]{Xiao-Liang Qi\thanks{xlqi@stanford.edu}}

\affil[1]{Department of Materials Science and Engineering, Stanford University, Stanford, CA 94305, USA}
\affil[2]{Department of Physics, Pennsylvania State University, University Park, PA 16802, USA}
\affil[3]{Department of Condensed Matter Physics, Weizmann Institute of Science, Rehovot 7610001, Israel}
\affil[4]{Path Integral Technology, Inc., Belmont, CA 94002, USA}

\date{September 2025}

\begin{document}

\maketitle

\begin{abstract}
We introduce QMBench, a comprehensive benchmark designed to evaluate the capability of large language model agents in quantum materials research. This specialized benchmark assesses the model's ability to apply condensed matter physics knowledge and computational techniques such as density functional theory to solve research problems in quantum materials science. QMBench encompasses different domains of the quantum material research, including structural properties, electronic properties, thermodynamic and other properties, symmetry principle and computational methodologies. By providing a standardized evaluation framework, QMBench aims to accelerate the development of an AI scientist capable of making creative contributions to quantum materials research. We expect QMBench to be developed and constantly improved by the research community.
\end{abstract}

\section{Introduction}

Benchmarks are essential for measuring and guiding the development of machine learning models. The rapid progress of large language models (LLMs) in recent years have been usually measured by benchmarks on common tasks such as math, coding, language processing etc.\cite{srivastava2023beyond,liang2022holistic,wang2024mmlu,jimenez2023swe} Specialized benchmarks such as GPQA\cite{rein2024gpqa} in have also been developed, but mainly at the level of course exam problems. Among the potential applications of LLM, scientific research is one of the most exciting areas that is being explored. If LLM can execute research tasks with high enough accuracy, it will be able to play the role of a research assistant or even a research collaborator. More and more attempts have been made to build AI scientist agents in multiple fields including biology\cite{huang2024crispr,huang2025biomni,ghareeb2025robin}, chemistry\cite{darvish2025organa,zhang2025multimodal} and more generic tasks\cite{gottweis2025towards,novikov2025alphaevolve,chai2025scimaster,lu2024ai}. However, a broader application of LLM in frontier scientific research remains challenging. A main difficulty is that research-level benchmarks are still scarce. For the purpose of evaluating AI's performance in frontier research tasks, research benchmarks need to satisfy the following criteria:
\begin{enumerate}
    \item {\bf Proper granularity}. Benchmarks covering a broad area such as physics and math are too broad for evaluating AI's research capability in a particular subfield. The scope of a benchmark needs to be defined by the experts in the field, similar to how the scope of an academic conference is defined. 
    \item {\bf Research-level questions}. Even in domains where AI performs well on graduate level exam problems, substantial differences remain between these tasks and actual research problems encountered by researchers. Identifying the truly research level problems require the problems to be selected by domain experts.
    \item {\bf Coevolving with the field}. Research problems are constantly evolving. New discoveries are made and new concepts are being introduced. The benchmarks need to be regularly updated to reflect the current knowledge and interest of each subfield. Such a mechanism has not been developed yet.
\end{enumerate}

Quantum materials constitute a broad class of solids in which quantum-mechanical effects at the level of electrons, spins, and lattice degrees of freedom give rise to qualitatively new phases of matter and functionalities. Prototypical examples include topological insulators and semimetals~\cite{hasan2010colloquium,qi2011topological,armitage2018weyl}, low-dimensional materials~\cite{andrei2021marvels,novoselov20162d}, unconventional superconductors~\cite{stewart2011superconductivity, scalapino2012common}, and strongly correlated oxides~\cite{imada1998metal, chakhalian2014colloquium}, where band structure, electronic correlations, and symmetry intertwine in nontrivial ways. Progress in this field requires deep expertise in quantum mechanics and solid-state physics, together with a diverse skill set: the ability to perform analytical derivations and approximate calculations, a firm grasp of crystal symmetry, group theory, and related mathematical structures, proficiency with first-principles electronic-structure simulation methods such as density functional theory (DFT) and beyond, and an understanding of the numerical algorithms and high-performance computing that underlie modern simulations. These features make quantum materials research an especially demanding yet attractive testbed for evaluating the emerging capabilities of AI assistants and agents.

Motivated by the need for research-level benchmarks, in this paper, we introduce QMBench, a benchmark set for quantum material research. In particular, we focus on crystalline materials. The problems in this benchmark cover different aspects of this field, classified by physical properties and research methods. In addition to fundamental principles and physical properties, we put an emphasis on density functional theory, since it plays a critical role in theoretical understanding of solid state materials. Our benchmark uncovers the strength and weakness of current LLMs in this field, which we discuss in more detail later in the draft. 

The remainder of the paper is organized as follows. In Sec. \ref{sec:description} we describe the problems in QMBench and provide example problems. In Sec. \ref{sec:performance} we summarize the performance of the leading models, including the scores and detailed analysis of some example problems. In Sec. \ref{sec:related works} we discuss other related works, and finally we conclude in Sec. \ref{sec:conclusion}. Our benchmark is developed and posted on \url{https://bench.science}, an open platform to facilitate collaboration and sharing of scientific research benchmarks.

\section{Detailed description of QMBench}\label{sec:description}

\subsection{Problem Statistics}
Quantum materials research constitutes an inherently multimodal endeavor that integrates theoretical derivation, computational implementation, and the interpretation of simulation data. To comprehensively evaluate these capabilities, our benchmark dataset addresses the entire research workflow, ranging from conceptual formulation to practical execution. As detailed in Table \ref{tab:answer_type_distribution}, the tasks require four distinct output modalities: multiple-choice selections, numerical values, free-text responses, and atomic structure files in the POSCAR format \cite{kresse1996efficiency}, a standard within the DFT community. The dataset comprises 103 problems organized into five thematic domains: structural properties, symmetry principles, computational methodologies, electronic properties, and thermal, optical, and magnetic properties. Table \ref{tab:difficulty_distribution} illustrates the distribution across these categories, which contain 14, 28, 18, 43, and 16 problems, respectively, with multi-label tagging allowing a single problem to span multiple domains. Furthermore, we stratify the problems into three difficulty tiers corresponding to undergraduate-level fundamentals, graduate-level knowledge, and frontier research challenges.

\begin{table}[htbp]
\centering
\footnotesize
\caption{Distribution of problems by answer type.}
\begin{tabular}{lcccc}
\toprule
\textbf{Answer Type} & \textbf{Multi-Choice} & \textbf{Numerics} & \textbf{POSCAR} & \textbf{Text} \\
\midrule
Number of Questions & 50 & 18 & 4 & 31 \\
\bottomrule
\end{tabular}
\label{tab:answer_type_distribution}
\end{table}

\begin{table}[htbp]
\centering
\caption{Distribution of problems by topic and difficulty level.}
\small 
\resizebox{\textwidth}{!}{\begin{tabular}{lccccc}
\toprule
\textbf{Difficulty} & \makecell{\textbf{Structural}\\\textbf{Properties}} & \makecell{\textbf{Symmetry}\\\textbf{Principles}} &
\makecell{\textbf{Computational}\\\textbf{Methodologies}} &
\makecell{\textbf{Electronic}\\\textbf{Properties}} &
\makecell{\textbf{Thermal, Magnetic}\\\textbf{and Optical Properties}} \\
\midrule
Easy   & 0  & 7  & 4  & 4  & 6  \\
Medium & 6  & 6  & 2  & 7  & 7  \\
Hard   & 8  & 15 & 12 & 32 & 3  \\
\midrule
Total  & 14 & 28 & 18 & 43 & 16 \\
\bottomrule
\end{tabular}}
\label{tab:difficulty_distribution}
\end{table}

\subsubsection{Structural Properties}
This category assesses the AI scientist’s understanding of atomic structures. In quantum materials research, constructing appropriate atomic configurations for first-principles calculations and interpreting structural features revealed by computational results are essential. Accordingly, the tasks include generating complex atomic structures, evaluating atomic forces, and determining the mechanical stability of structures from simulation outcomes.

\begin{tcolorbox}[ colframe=blue!60!yellow, colback=blue!5!white, title={Example Problem}]
\textbf{Heterostructure supercell construction}

\vspace{0.1in}
I plan to study a heterostructure between graphene and monolayer CrSBr. They do not match in the lattice parameters. What is the optimal supercell I can choose in DFT calculations? Please specify it in the unit of CrSBr lattice vectors.

\vspace{0.1in}
\textbf{Problem Type}: Text
\end{tcolorbox}

\subsubsection{Symmetry Principles}
This category evaluates the AI scientist’s knowledge of crystal symmetry and its implications for material properties. Problems cover the theory of space groups and their representations, as well as the knowledge of crystal-symmetric constraints on materials' band structures and response functions.

\example{Symmetry of diamond}
{What symmetry operations does the space group of diamond contain? Select all that apply.

A. $C_3$ (three-fold rotation)

B. $C_6$ (six-fold rotation)

C. ${M_y| t=(1/6,0,1/6)}$ (glide reflection symmetry)

D. $M$ (mirror symmetry)

E. $C_4$ (four-fold rotation)

F. $P$ (inversion symmetry)
}{Multiple Choice}

\subsubsection{Computational Methodologies}
This category focuses on the AI scientist’s ability to employ first-principles computational methods and to select appropriate parameters. Problems include generating formatted structural files [e.g. POSCAR used by a popular density-functional theory(DFT) package] , configuring DFT parameters, and choosing exchange–correlation functionals and other computational settings.

\example{POSCAR for $\text{Bi}_2 \text{Te}_3$}
{Output the atomic positions of a slab of five van der Waals layers of $\text{Bi}_2 \text{Te}_3$ in POSCAR format for DFT calculations. Please set the thickness of vacuum layer to be 15 angstroms. Your output should be the content of the POSCAR file, in a string.}{POSCAR}

\subsubsection{Electronic Properties}
This category examines the AI scientist’s capacity to understand and study the electronic structure of materials. It covers theoretical derivations within band theory, including band topology, first-principles calculations of electronic structures, and knowledge of fundamental facts related to electronic structure.

\example{Band gap estimation from figure}
{Please provide an estimate of the band gap of this material in electronvolts (eV). Only give the numerical value, without including the unit.
\begin{center}
    \includegraphics[width=0.5\textwidth]{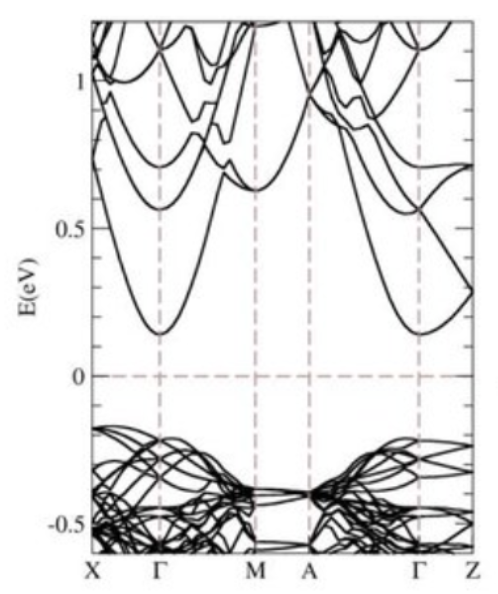}
\end{center}
}{Numerics}

\example{Symmetry constraint of nearest neighbor hopping in tight-binding model} 
{Consider atoms on a two-dimensional triangular lattice with the lattice constant $a$ and basis vectors $\mathbf{a}_1 = (a,0), \mathbf{a}_2 = (a/2,\sqrt{3}a/2)$. The system has $D_{3h}$ symmetry and time-reversal symmetry. On each site, consider the 3 $d$-orbitals $d_{z^2}$, $d_{xy}$ and $d_{x^2-y^2}$ and neglect spin. The lattice has $D_{3h}$ symmetry group with mirror symmetry along xy and yz planes. Define the nearest neighbor hopping matrix elements along $\mathbf{a}_1$ direction as $t^{(1)}_{ij} = \langle \mathbf{a}_1,i | H | \mathbf{0},j \rangle$ with $i,j=1,2,3$ corresponding to $d_{z^2}, d_{xy}, d_{x^2-y^2}$ respectively. Similarly, define the nearest neighbor hopping matrix element along $\mathbf{a}_2$ direction as $t^{(2)}_{ij}$. 

Select the correct expression of $t^{(2)}_{22}$ and $t^{(2)}_{33}$ in terms of $t^{(1)}_{22}$ and $t^{(1)}_{33}$.

A. $t^{(2)}_{22} = t^{(1)}_{22}$, $t^{(2)}_{33} = t^{(1)}_{33}$

B. $t^{(2)}_{22} = t^{(1)}_{22}$, $t^{(2)}_{33} = -t^{(1)}_{33}$

C. $t^{(2)}_{22} = \frac{1}{2} t^{(1)}_{22} + \frac{\sqrt{3}}{2} t^{(1)}_{33}$, $t^{(2)}_{33} = -\frac{\sqrt{3}}{2} t^{(1)}_{22} + \frac{1}{2} t^{(1)}_{33}$

D. $t^{(2)}_{22} = \frac{1}{4} t^{(1)}_{22} + \frac{3}{4} t^{(1)}_{33}$, $t^{(2)}_{33} = \frac{1}{4} t^{(1)}_{22} + \frac{3}{4} t^{(1)}_{33}$

E. $t^{(2)}_{22} = \frac{1}{4} t^{(1)}_{22} + \frac{3}{4} t^{(1)}_{33}$, $t^{(2)}_{33} = \frac{3}{4} t^{(1)}_{22} + \frac{1}{4} t^{(1)}_{33}$ 
}{Multiple Choice}

\subsubsection{Thermal, Optical and Magnetic Properties}
This category evaluates the AI scientist’s ability to study how materials respond to external conditions such as temperature, magnetic fields, and light, encompassing a range of thermal, optical, and magnetic phenomena.

\example{Dirac semimetal specific heat} 
{The specific heat contribution from electrons in a 3D Dirac semimetal at low temperature  (much lower than Fermi temperature) is proportional to $T^n$, where $T$ is the temperature. Please give the power $n$.}
{Numerics}

\subsection{Evaluation Method}
To automatically evaluate the AI scientist (student model) across diverse problems in quantum materials, we define several standardized problem types, and define grading algorithm based on each type. In more details, we have the following problem types:

\noindent{\bf Multiple-choice}. The answer is extracted and compared with the ground truth. When the ground truth includes multiple answers, an incomplete answer receives partial credit, but any wrong answer receives zero credit. For example, if the ground truth is $A,B,D$, answer $A,B$ will receive $2/3$ of the full score, while $A,C$ will receive zero.

\noindent{\bf Numerics}. The answer is an integer, or a floating-point number (with error bar). For integer, credit is given only for exact matching. For floating-point number, the ground truth is a range (such as $[0.1,0.12]$) and the answer is considered correct if it falls in this range. 

\noindent{\bf Text}. For questions with a more free-form answer, we allow the answer to be text, which will be graded by another LLM agent. To minimize subjectivity in LLM grading, each problem includes a detailed rubrics to explicitly enumerate the required points and their corresponding scores, which provides an objective and reproducible basis for evaluation.

\noindent{\bf POSCAR}. Some problems require the model to output a POSCAR file, which is a commonly used format of crystal structure data. The POSCAR is compared with the ground truth with a symmetry-aware grading function. Different atom coordinates could correspond to the same structure since they may be related by a translation or a rotation. Our grading function considers such coordinate transformation ambiguity and also allows an error tolerance. The tolerance for lattice angles is 1° and that for cell lengths is 5\%. The function generates candidate rotation/reflection matrices and translation vectors to seek for a possible matching. 
By relying on this symmetry-aware comparison, we can robustly handle equivalent structures that are presented in different cell orientations or with different origins. Full credit is awarded if the two structures differ only by a rigid-body translation or rotation, and their lattice constants and atomic coordinates lie within the relative error margin.

\section{Model Performance}\label{sec:performance}

\subsection{Statistics}

\begin{figure}[htbp]
    \centering
    \includegraphics[width=1\textwidth]{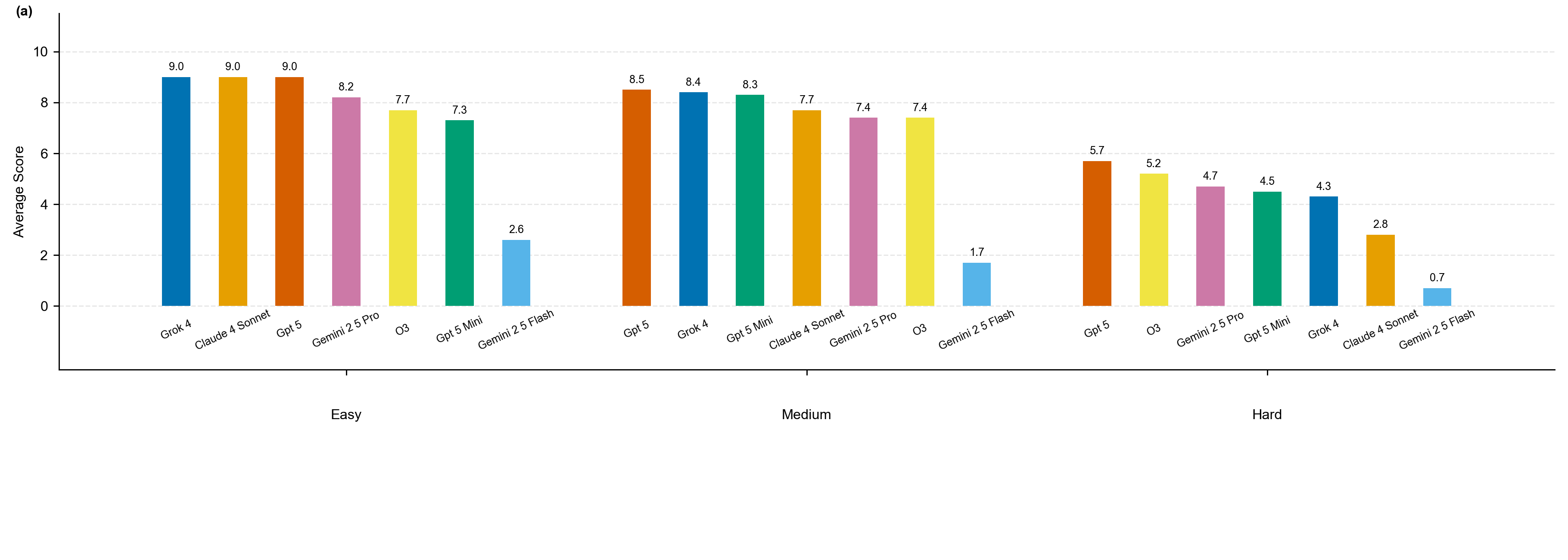}
    \includegraphics[width=1\textwidth]{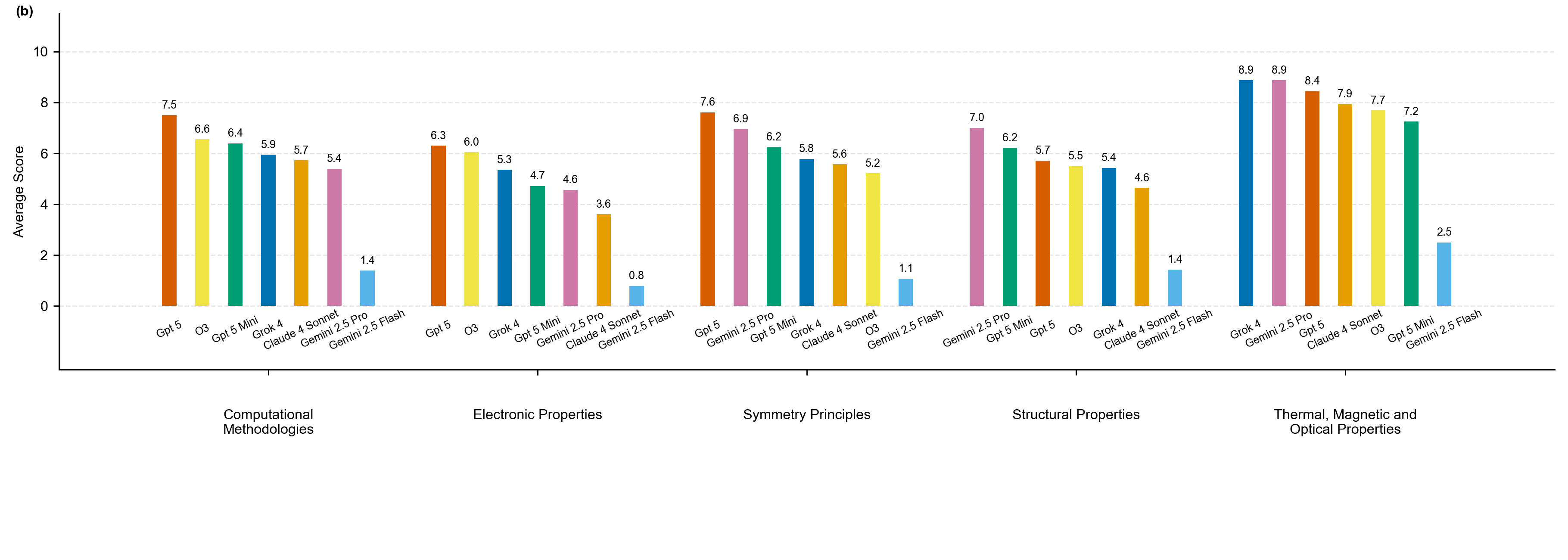}
    \caption{Performance of multimodal LLMs on QMBench for questions at different difficulty levels. (a) Performance on questions at different difficulty levels. (b) Performance on questions of different categories.}
\end{figure}

We evaluate two groups of LLMs on our benchmark: text-only models, which are tested on the 92 purely textual problems, and multimodal models, which are tested on the full set of 103 problems including 11 requiring visual interpretation. All problems are scored out of 10 points.

For the text-only setting, five representative open-source and commercial models were evaluated: DeepSeek v3.1, DeepSeek-R1, Kimi-K2, Qwen-3, and GPT-OSS-120B. Overall accuracy remained low, with model averages ranging from 4.2 to 5.1 points. The strongest performer was DeepSeek-R1 (5.05), followed by GPT-OSS-120B (4.80), Qwen-3 (4.75), DeepSeek v3.1 (4.47), and Kimi-K2 (4.17).

For the multimodal benchmark, which includes problems requiring interpretation of figures from computational and experimental work, the participating models were Grok-4, Claude 4 Sonnet, GPT-5-mini, Gemini 2.5 Pro, Gemini 2.5 Flash, GPT-5, and O3. Performance was somewhat higher at the top end, with GPT-5 achieving the highest overall average (7.30), followed by O3 (6.41), Gemini 2.5 Pro (6.27), GPT-5-mini (6.21), and Grok-4 (6.53). Claude 4 Sonnet averaged 5.53, while Gemini 2.5 Flash lagged significantly at only 1.39. Since 92 of the 103 questions are purely textual, these performance differences should be attributed primarily to stronger language and reasoning capabilities rather than to superior image understanding.

Across categories, several consistent trends emerge. Thermal, Optical, and Magnetic Properties were relatively the easiest for multimodal models, with average scores exceeding 7.0 and top models above 8.0. Electronic Properties posed significant difficulty for all models, with text-only systems averaging below 3.0, while multimodal systems achieved moderate gains (average 4.48, with GPT-5 reaching 6.30). Computational Methodologies and Symmetry Principles fell in the intermediate range, with averages around 4.2–5.0 for text-only models and 5.5–5.7 for multimodal models. Structural Properties remained challenging, with scores clustering near 4.2 in the text-only group and 5.1–5.4 in the multimodal group.

Taken together, these results demonstrate that our benchmark is difficult for current AI systems. Even frontier multimodal models such as GPT-5 and O3 fail to exceed 7.5/10 in most categories, and text-only models rarely reach 5/10. Persistent weaknesses are evident in electronic structure reasoning, computational setup, and symmetry analysis, while the ability to handle physical responses to external fields (thermal, optical, magnetic) remains uneven across models. These findings highlight the benchmark’s discriminative power and point to clear research challenges in building AI scientists capable of robust quantum materials research.

\subsection{Further analysis to model performance}
A detailed analysis of model performance on QMBench reveals a sharp dichotomy in capabilities. The leading models performed exceptionally well on knowledge-oriented questions, functioning as highly effective knowledge resources. Across categories, they demonstrated strong performance on items requiring the identification of standard terminology, the summarization of textbook-level relationships, or the recall of canonical examples (e.g., prototypical topological materials, or the common choice of exchange–correlation functionals).

However, this proficiency in conceptual recall stands in stark contrast to their performance in tasks requiring applied reasoning and practical execution. Beyond aggregate scores, a closer inspection of error patterns reveals several systematic limitations.

First, problems requiring rigorous analytical calculations and derivations proved exceptionally challenging. Even the best-performing model (GPT-5) collectively answered 13 such questions incorrectly. These failures often reflected a fundamental inability to reliably apply group-theoretical arguments or execute multi-step algebraic manipulations, even when the relevant foundational concepts had been correctly articulated in earlier parts of the response.

Second, despite nominal multimodal capabilities, current models faltered on questions requiring quantitative figure interpretation. Errors persisted even on fundamental tasks, such as accurately enumerating the number of bands crossing the Fermi level. This suggests that current LLM-based agents struggle with the meticulous visual inspection and figure-based summarization that are central to interpreting computational and experimental results in quantum materials research.

Third, regarding questions involving atomistic structures, the models demonstrated an incomplete command of structural representations. They showed reasonable familiarity with standard formats such as POSCAR files: for example, among the four tasks that require the generation or modification of POSCAR, GPT-5 solved two correctly. However, all models failed on more complex structural tasks, such as slab geometry construction, which demands consistent handling of surface terminations and vacuum regions. We expect such problems are useful tests for the effectiveness of external tools, which can be provided to the LLM to carry specialized computations. In the current evaluation we haven't included such tools. 


Our findings collectively highlight this distinct performance profile. LLMs already function as effective knowledge resources for quantum materials science, but substantial gaps persist in their ability to perform sustained quantitative reasoning, rigorous derivations, precise figure interpretation, and robust programming workflows. In the following we will carry some more detailed analysis on LLM's answers for some example problems.

\analysis
{Heterostructure supercell construction}
{I plan to study a heterostructure between graphene and monolayer CrSBr. They do not match in the lattice parameters. What is the optimal supercell I can choose in DFT calculations? Please specify it in the unit of CrSBr lattice vectors.}
{
\vspace{1pt}

1. Identify the lattice constant of graphene and CrSBr.

2. Recognizing different shapes of graphene and CrSBr's unit cell.

3. Calculating lattice constant ratios and supercell matching condition.

\vspace{-3pt}

\begin{center}
\begin{tabular}{|l|c|c|c|c|c|}
\hline
\textbf{Model} & \textbf{$a_{\mathrm{CrSBr}}$} & \textbf{$b_{\mathrm{CrSBr}}$} & Step 1 & Step 2 & Step 3 \\ \hline
Grok 4 & 3.48 & 4.79 & \checkmark & \checkmark & \checkmark \\ \hline
Gemini 2.5 Pro & 3.51 & 4.77 & \checkmark & \checkmark & \checkmark \\ \hline
GPT O3 & 3.47 & 4.79 & \checkmark & \checkmark & \halfcheckmark \\ \hline
GPT 5 & 3.50 & 4.79 & \checkmark & \checkmark & \halfcheckmark \\ \hline
DeepSeek V3.1 & 3.50 & 4.84 & \checkmark & \checkmark & \halfcheckmark \\ \hline
Claude 4 Sonnet & 3.48 & 4.85 & \checkmark & \checkmark & \cross \\ \hline
Gemini 2.5 Flash & 3.49 & 4.70 & \checkmark & \checkmark & \cross \\ \hline
DeepSeek R1 & 3.50 & 4.70 & \checkmark & \cross & \cross \\ \hline
Kimi K2 & 3.58 & 4.80 & \checkmark & \cross & \cross \\ \hline
GPT 5 mini & \multicolumn{2}{c|}{ - (no data)} & \cross & \cross & \cross \\ \hline
Qwen 3 & \multicolumn{2}{c|}{3.00 (hexagonal, \cross)} & \cross & \cross & \cross\\ \hline
GPT OSS 120B & 3.55  & 4.05 (\cross) & \cross & \cross & \cross\\ \hline
\end{tabular}
\end{center}
\vspace{-10pt}

}{Almost all LLMs correctly obtained the lattice constants of graphene and CrSBr within approximately 3\% accuracy, which falls within the typical tolerance range of DFT calculations, except for \textbf{GPT OSS 120B}, \textbf{GPT 5 mini}, and \textbf{Qwen 3}.

The second step involves recognizing that graphene possesses a hexagonal lattice, whereas CrSBr has a rectangular one. This requires artificial expanding graphene’s hexagonal unit cell into an equivalent rectangular cell of dimensions $a_{\text{graphene}} \times \sqrt{3}a_{\text{graphene}}$, allowing proper comparison with CrSBr’s rectangular lattice. At this stage, a few additional models, such as \textbf{DeepSeek R1}, \textbf{Kimi K2}, failed to make this correction.

The third step is identifying the correct superlattice matching condition. While several models (e.g., \textbf{Claude 4 Sonnet} and \textbf{Gemini 2.5 Flash}) provided only vague or incomplete reasoning, others, such as \textbf{GPT O3}, \textbf{GPT 5}, and \textbf{DeepSeek V3}, performed better by deriving near-accurate lattice matching relations. However, these models failed to take into account two distinct ways to align rectangular unit cells, e.g. $a_{\text{CrSBr}} \parallel a_{\text{graphene}}$ and $b_{\text{CrSBr}} \parallel a_{\text{graphene}}$. Only \textbf{Gemini 2.5 Pro} and \textbf{Grok 4} correctly considered both configurations and arrived at the optimum lattice-matching ratio.}

\analysis{Fermi surface counting}{
Here is the Fermi surface of a material. Can you tell me how many bands cross the Fermi energy (i.e., how many Fermi pockets exist)?
\begin{center}
    \includegraphics[width=0.5\textwidth]{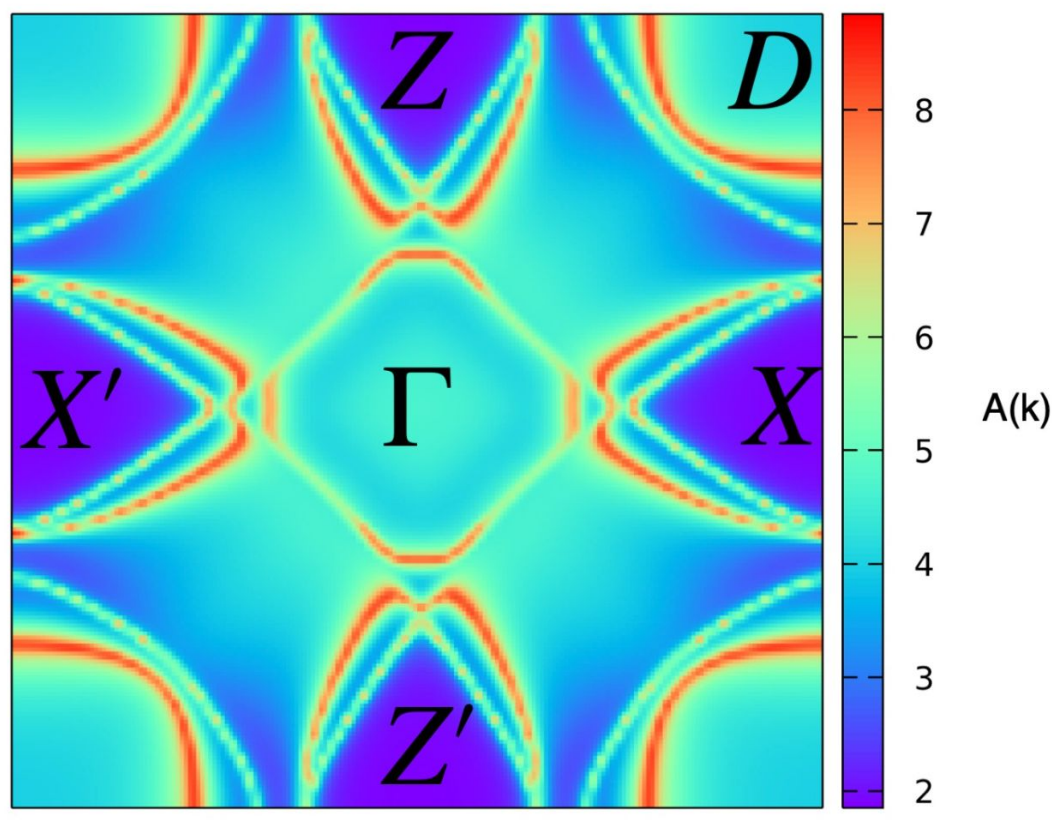}
\end{center}
}{

\vspace{-10pt}
\begin{center}
\begin{tabular}{@{}p{0.85\linewidth}c@{}}
From \textbf{Claude 4 Sonnet}: & \\
\textit{- 1 central pocket at $\Gamma$} & \checkmark \\
\textit{- 4 corner pockets (at the X and equivalent points)} & \cross \\
\textit{- 2 additional pockets in the Z regions} & \checkmark \\
From \textbf{Gemini 2.5 Pro}: & \\
\textit{* From the $\Gamma$ point: 1 pocket} & \checkmark \\
\textit{* From the corner (D) points: 1 pocket} & \cross \\
\textit{* From the edge-center (X, Z, X', Z') points: 8 pockets} & \cross \\
From \textbf{Gemini 2.5 Flash}: & \\
\textit{* 1 Central Pocket} & \checkmark \\
\textit{* 4 Intermediate (Lens-shaped) Pockets} & \checkmark \\
\textit{* 4 Corner Pockets} & \cross \\
\end{tabular}
\end{center}

\vspace{-15pt}

}
{
All models correctly identified the single Fermi pocket around the $\Gamma$ point ($n_{\Gamma} = 1$).
However, most models miscounted pockets along the Brillouin zone (BZ) edges—where periodic boundaries must be patched across X(X'), Z(Z'), and D corners—which leads to their failure to obtain the correct answer $n_X = 2$, $n_Z = 2$, and $n_D = 2$.
Typical errors include: \textbf{Claude 4 Sonnet} ($n_X = 4$, $n_D = 0$), \textbf{Gemini 2.5 Pro} ($n_X = n_Z = 4$, $n_D = 1$), and \textbf{Gemini 2.5 Flash} ($n_D = 4$).

\vspace{-7pt}

\begin{center}
\begin{tabular}{@{}>{\centering\arraybackslash}p{0.25\textwidth}@{}
                >{\centering\arraybackslash}p{0.25\textwidth}@{}
                >{\centering\arraybackslash}p{0.25\textwidth}@{}
                >{\centering\arraybackslash}p{0.25\textwidth}@{}}
    \includegraphics[width=0.95\linewidth]{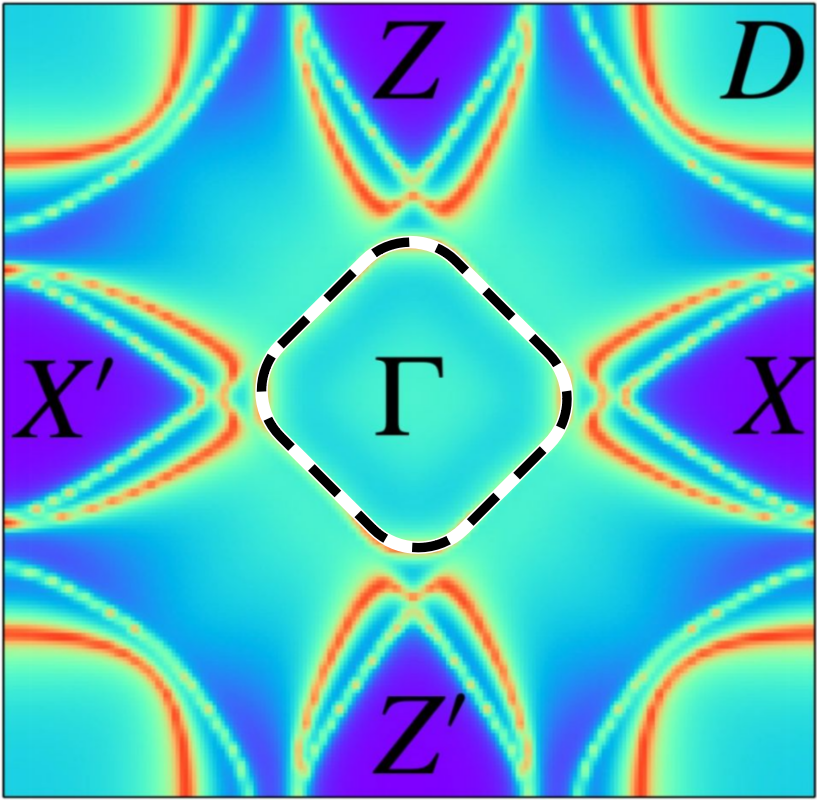} &
    \includegraphics[width=0.95\linewidth]{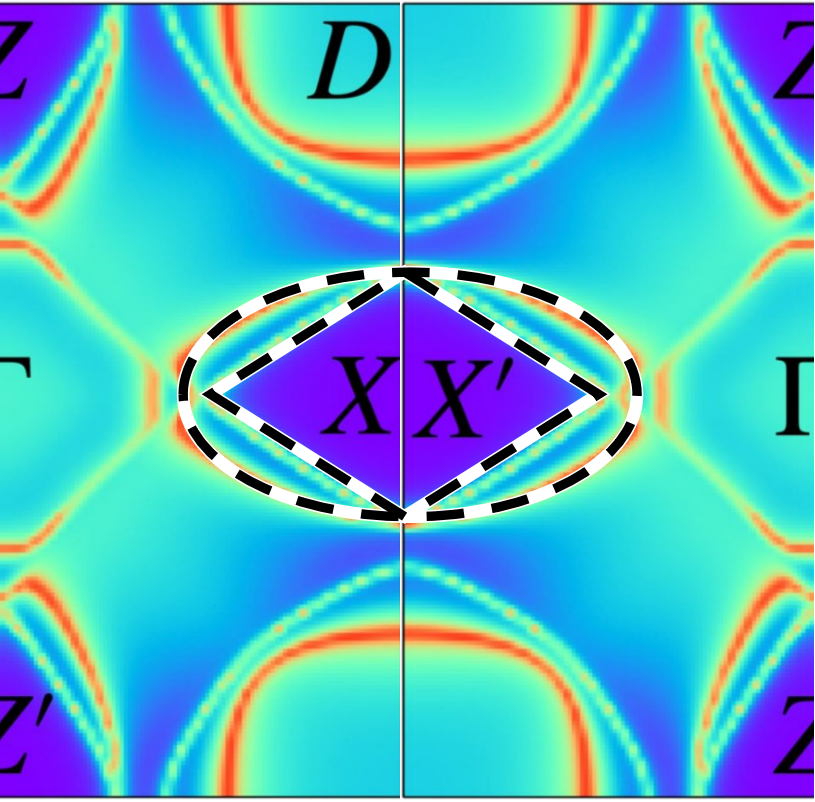} &
    \includegraphics[width=0.95\linewidth]{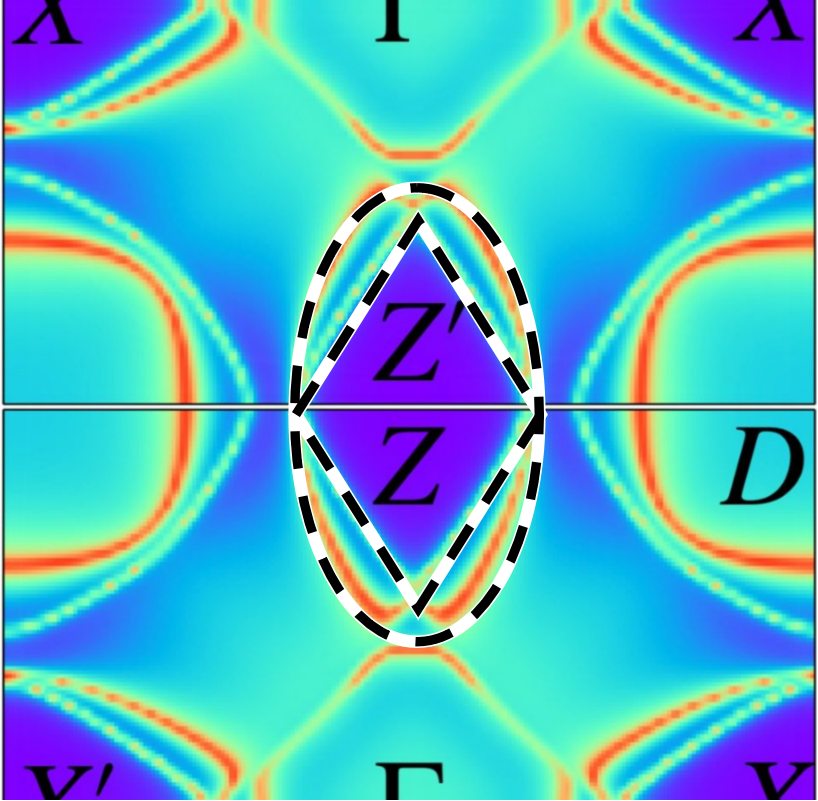} &
    \includegraphics[width=0.95\linewidth]{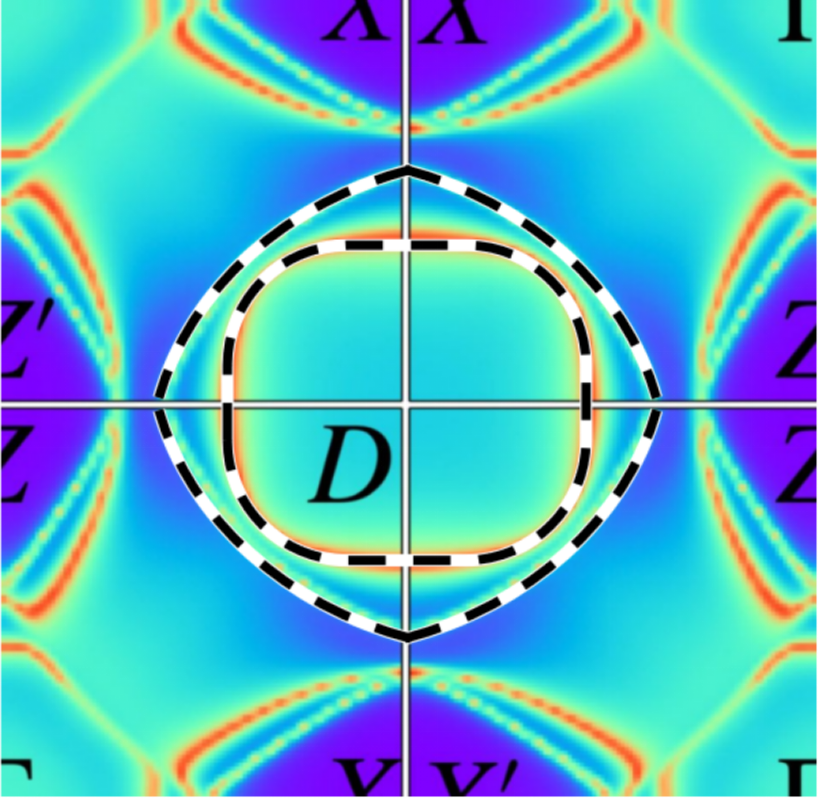} \\
    $n_{\Gamma} = 1$ &
    $n_X = 2$ &
    $n_Z = 2$ &
    $n_D = 2$ 
\end{tabular}
\end{center}

\vspace{-15pt}

}

\analysis{Number of parameters in tight-binding model}
{
Consider atoms on a two-dimensional triangular lattice with the lattice constant $a$ and basis vectors $\mathbf{a}_1 = (a,0), \mathbf{a}_2 = (a/2,\sqrt{3}a/2)$. The system has $D_{3h}$ symmetry and time-reversal symmetry. On each site, consider the 3 $d$-orbitals $d_{z^2}$, $d_{xy}$ and $d_{x^2-y^2}$ and neglect spin. The lattice has $D_{3h}$ symmetry group with mirror symmetry along xy and yz planes. Define the nearest neighbor hopping matrix elements along $\mathbf{a}_1$ direction as $t^{(1)}_{ij} = \langle \mathbf{a}_1,i | H | \mathbf{0},j \rangle$ with $i,j=1,2,3$ the orbital index.

How many independent parameters are there in the matrix elements $t^{(1)}_{ij}$? Count the number of real-valued parameters (so that one complex number counts as two real numbers). 

\vspace{2pt}

(A) 3 \qquad (B) 5 \qquad (C) 6 \qquad (D) 9
}
{

\vspace{1pt}

1. Identify the $3\times 3$ hopping matrix elements

\vspace{2pt}

\begin{tabular}{|p{0.96\linewidth}|}
\hline
\textbf{Gemini 2.5 Pro (\checkmark)} \\ \hline
\textit{$T \equiv t^{(1)}_{ij} = \langle \mathbf{a}_1, i | H | \mathbf{0}, j \rangle$ is a $3 \times 3$ matrix. Without any symmetry, it would have 9 complex elements, which amounts to 18 real parameters.} \\ \hline
\end{tabular}

\vspace{2pt}

2. Implement time-reversal symmetry ($\mathcal{T}$) constraint

\vspace{2pt}

\begin{tabular}{|p{0.60\linewidth}|p{0.32\linewidth}|}
\hline
\textbf{Claude 4 Sonnet (\checkmark)} &
\textbf{DeepSeek V3 (\cross)} \\ \hline
\textit{$\mathcal{T}$ requires the Hamiltonian to be real. This means all matrix elements must be real. So we start with 9 real parameters.} &
\textit{Under $\mathcal{T}$, the hopping matrix must satisfy \( t_{ij} = t_{ji}^* \).} \\ \hline
\end{tabular}

\vspace{2pt}

3. Identify the representation of $\mathcal{M}_{yz}$ mirror symmetry

\vspace{2pt}

\begin{tabular}{|p{0.96\linewidth}|}
\hline
\textbf{Gemini 2.5 Pro (\checkmark)} \\ \hline
\textit{The representation matrix of $\mathcal{M}_{yz}$ is: }$D(\mathcal{M}_{yz}) = $ \textit{diag(1,-1,1).} \\ \hline
\end{tabular}

\vspace{2pt}

4. Implement spatial symmetry ($\sigma_{v,yz}$ mirror) constraint

\vspace{2pt}

\begin{tabular}{|p{0.60\linewidth}|p{0.32\linewidth}|}
\hline
\textbf{Gemini 2.5 Pro (\checkmark)} &
\textbf{Kimi K2 (\cross)} \\ \hline
\textit{
For an operation $g$ that relates the hopping vector $\mathbf{R}$ to $g\mathbf{R}$, the constraint is:
$t^{(\mathbf{R})} = D(g)^* t^{(g\mathbf{R})} D(g)$. The hopping matrix to the site $-\mathbf{a}_1$ is related to the hermitian conjugate of the hopping matrix to $\mathbf{a}_1$: $t^{(-\mathbf{a}_1)} = (t^{(\mathbf{a}_1)})^\dagger = T^\dagger$. The spatial symmetry constraint simplifies to:
$T = D T^T D$} &
\textit{So $t^{(1)}$ must satisfy:
$t^{(1)'}_{ij} = D_{ii'} t^{(1)}_{i'j'} D_{j'j}^{-1} = t^{(1)}_{ij}$
This implies that $t^{(1)}_{12} = -t^{(1)}_{12}$ (so $t^{(1)}_{12} = 0$), $t^{(1)}_{21} = -t^{(1)}_{21}$, etc.} \\ \hline
\end{tabular}  

}{Almost all LLMs can identify the hopping matrix to be $3 \times 3$, as well as representation matrix of $\mathcal{M}_{yz}$ to be diag(1,-1,1). However, when implementing $\mathcal{T}$ and $\mathcal{M}_{yz}$, some models, including \textbf{DeepSeek V3} and \textbf{Kimi K2}, didn't correctly consider the transformation of marix elements upon the flip of hopping direction $\mathbf{a}_1\rightarrow -\mathbf{a}_1$ under $\mathcal{T}$ or $\mathcal{M}_{yz}$.}

\analysis{Landauer-Buttiker transmission matrix}
{We consider a two-band tight-binding Hamiltonian on a square lattice,
$$
H_0 = (m_0 + 4m_2)s_z, 
\\
H_x = -\frac{i}{2}A s_x - m_2 s_z, 
\\
H_y = -\frac{i}{2}A s_y - m_2 s_z,
$$
where $H_0$ is the onsite energy, $H_x, H_y$ are respectively the nearest-neighbor hoppings in the positive $x$ and $y$ directions. Here $s_{x,y,z}$ label the Pauli matrices. The sample is a square grid with dimensions $L$. Each lead has the width $L$, in full contact with each side of the sample and shares the same Hamiltonian with the sample.

Take the sample size $L=50$. The model parameters are $A=0.5, m_0=0.3, m_2=-1$. Consider the Fermi energy $\mu=0.5$.

Please calculate the transmission matrix among the four leads according to the Landauer-Buttiker formalism, at chemical potential $\mu$, assuming zero temperature. Here we order the leads clockwise, starting from the lead in the negative-$x$ direction, and the basis for the transmission matrix is arranged in the same order. Please keep the diagonal terms of the transmission matrix, which represents the reflection probability at each lead.}
{from \textbf{Gemini2.5 Pro}'s -
\textit{
The given Hamiltonian is a lattice realization of a QAHI. We analyze its bulk properties using the parameters $A=0.5, m_0=0.3, m_2=-1$. The energy spectrum of the bulk has a gap. Although the given Fermi energy $\mu=0.5$ appears to be in the bulk conduction band based on a simple analysis at the $\Gamma$-point, problems of this type on topological insulators are almost always designed to probe the physics of edge states, which requires the Fermi energy to be within the bulk gap. We proceed under the physically reasonable assumption that the parameters are intended to place the system in the topological regime with the Fermi energy inside the gap. In this regime, electron transport occurs exclusively through topologically protected edge states.
}
}
{Even though \textbf{Gemini 2.5 Pro} realizes the system is in a metallic phase, it tends to reinterpret or revise the problem into a more familiar form, possibly reflecting biases inherited from the types of problems seen during training. Therefore it proceeds as if the system is in a topological phase with Chern number $C=1$, leading to a simplified picture of perfect chiral transport with no backscattering. In contrast, the true transmission matrix elements in a metallic phase are not quantized, and need to be obtained from an explicit computation.
}

\section{Related Works}\label{sec:related works}

A widely used science benchmark is the Graduate-level Google-Proof Q\&A benchmark (GPQA)\cite{rein2024gpqa}, but it is mainly at the coursework rather than research level. SciCode\cite{tian2024scicode} is a benchmark focusing on coding  tasks in science. Humanity's Last Exam (HLE)\cite{phan2025humanity} is a set of difficult questions covering different fields of math and science. Several more field-specific benchmarks have been developed this year. The theoretical physics bench (\href{https://tpbench.org/}{TPBench}) \cite{chung2025theoretical} is a benchmark on theoretical physics, mainly in the areas of high energy physics and cosmology. Ref.~\cite{xu2025physense} introduced PhySense, a benchmark focusing on physics reasoning. Two recent works have introduced benchmarks in condensed matter physics\cite{wang2025cmphysbench,pan2025cmt}. Among them, Ref.~\cite{pan2025cmt} covered many numerical methods such as exact diagonalization, Monte Carlo, DMRG, Hartree-Fock, etc. Ref.~\cite{wang2025cmphysbench} developed an innovate approach to automatic evaluation based on Scalable Expression Edit Distance (SEED).

\section{Conclusion and Discussion}\label{sec:conclusion}

In summary, we have developed QMBench, a comprehensive benchmark designed to evaluate the research capabilities of large language models in quantum materials science.

Our findings reveal a distinct performance dichotomy. On one hand, current models function as effective encyclopedic resources, demonstrating strong performance on knowledge-oriented questions that test the recall of established concepts, definitions, and qualitative trends.

On the other hand, we identify substantial gaps in tasks requiring applied reasoning and practical execution. These limitations are systematic and include:

1. A failure to perform rigorous analytical derivations and multi-step algebraic manipulations.

2. Poor performance in the quantitative interpretation of figures, such as band-structure plots, despite nominal multimodal capabilities.

3. An incomplete command of complex atomistic structural manipulations, particularly for non-trivial geometries like slab construction.

Collectively, these findings indicate that while LLMs have mastered the knowledge base of quantum materials, significant challenges remain in bridging the gap from conceptual recall to the robust, multi-faceted reasoning and practical application required for authentic scientific research.

We would like to make some further discussion on different ways to use a benchmark to probe an AI model. In our benchmark, we include problems that would be much simpler for a model with access to tools such as computational software or code execution. Although our current evaluation is carried on models without such capability, we can also apply the benchmark to AI agents that have these tools. In general, an AI agent is defined by three aspects: (1) the foundation model(s); (2) the prompts and the collaboration architecture (if there are multiple agents); (3) the tools. In general, if a benchmark contains questions that certain tools will be useful for, one can use the benchmark to independently evaluate these three parts. For example, we can fix the foundation model and compare the performance with and without different tools, which provides an evaluation of the usefulness of the tools. We can fix the tools and the architecture and switch the foundation model to evaluate the capability of the models. By comparing the capability of the models with and without tools, we can also evaluate the tool-use capability of the models in this family of tasks. We can also fix both the model and the tools, and test how prompt engineering or adjustment of the multi-agent collaboration pattern can affect the result. 

Finally, we would like to make some further comments about the platform \url{https://bench.science}. The goal of this platform is to facilitate the collaboration on benchmarks in scientific research. A group of researchers can collaborate by posting questions, evaluate the question against a list of models, test the grading by the grading model, and comment and approve on each other's questions. When the set of benchmarks is ready, it can be published, which provides a uniquely identifiable version of this benchmark. The published benchmark is not entirely public. The organizers of the benchmark can set certain problems to be public, and keep other problems private, to avoid data pollution. If the organizers want, they can also accept new questions submitted by the community. The goal is to have a "github" for benchmark collaborations. We welcome submissions of new questions in the field of quantum materials to our benchmark. You can submit your questions at \href{https://bench.science/bench/0f864f53-b7df-4525-bc37-1349e7aca763}{our project page}. 

{\bf Acknowledgement}. We would like to thank Chen Nie, Yang Si, Rong Tao, Bojun Yan from Path Integral Technology for providing engineering support. We would like to thank Eun-Ah Kim and Moritz M\"unchmeyer for helpful discussions. Y.W. and F.H.J. acknowledge support by the National Science Foundation (NSF) CAREER award through Grant No. DMR-2238328. B.Y. acknowledges the financial support by the Israel Science Foundation (ISF: 2974/23) and the Penn State Materials Research Science and Engineering Center for Nanoscale Science under National Science Foundation award DMR-2011839. H.T.L and C.X.L were partially supported by Rising Researcher award ICDS\_RR25\_27708 from Penn State's Institute for Computational \& Data Sciences (RRID:SCR\_025154).

\bibliographystyle{unsrt}
\bibliography{refs}
\end{document}